\documentclass[a4paper,10pt]{article}
\usepackage{hyperref}
\usepackage{amsmath,amsfonts,amssymb}

\def\BA{\begin{eqnarray}}
\def\EA{\end{eqnarray}}
\def\BAN{\begin{eqnarray*}}
\def\EAN{\end{eqnarray*}}
\let\NN=\nonumber
\def\sfrac#1#2{\mbox{$\frac{#1}{#2}$}}
\def\DDt{\sfrac{\rm d}{{\rm d} t}\,}

\usepackage{amssymb,amsmath}

\def\BE{\begin{equation}}
\def\EE{\end{equation}}

\def\BA{\begin{eqnarray}}
\def\EA{\end{eqnarray}}
\def\BAN{\begin{eqnarray*}}
\def\EAN{\end{eqnarray*}}
\let\NN=\nonumber
\def\sfrac#1#2{\mbox{$\frac{#1}{#2}$}}
\def\DDt{\sfrac{\rm d}{{\rm d} t}\,}

\title{On the generation of drift flows in wall-bounded flows\\
transiting to turbulence}

\author{Paul Manneville\\[1ex]
{\normalsize Hydrodynamics Laboratory, CNRS-UMR 7646,  \'Ecole Polytechnique, F-91128 Palaiseau, France}\\
{\normalsize \tt paul.manneville@ladhyx.polytechnique.fr}}
\date{\small \today\\
plain \LaTeX\ version of article to appear in Theoretical and Applied Mechanics Letters}
\begin{document}

\maketitle

\begin{abstract}
Despite recent progress, laminar-turbulent coexistence in transitional planar wall-bounded shear flows is still not well understood.
Contrasting with the processes by which chaotic flow inside turbulent patches is sustained at the local (minimal flow unit) scale, the mechanisms controlling the obliqueness of laminar-turbulent interfaces typically observed all along the coexistence range are still mysterious.
An extension of Waleffe's approach [Phys. Fluids 9 (1997) 883--900] is used to show that, already at the local scale, drift flows breaking the problem's spanwise symmetry are generated just by slightly detuning the modes involved in the self-sustainment process.
This opens perspectives for theorizing the formation of laminar-turbulent patterns.\\[1ex]
\end{abstract}

\section{Context and purpose}
Generically the transition to turbulence  in flows along solid walls, so-called wall-bounded flows, can be triggered at Reynolds numbers well below the value at which the laminar base flow profiles are linearly unstable \cite{Gr00,Ma16}.
The existence of nontrivial solutions to the Navier--Stokes equations (NSEs) competing with the stable base flow, is believed to be well understood in terms of self-regenerating coherent structures \cite{HKW95}.
The corresponding process, called SSP~\cite{HKW95,Wa97}, involves  {\it streamwise vortices\/} inducing {\it streamwise streaks\/} by lift-up, the so-produced mean flow distortion being subsequently unstable in a way that closes the cycle by feeding the vortices present at start.
At least at the moderate Reynolds numbers where the transition to turbulence takes place, this  plausible general sequence has a high degree of applicability.

Such nontrivial flow regimes arise from saddle-node bifurcations in state space but their coexistence with laminar base flow has also to be appreciated in physical space.
An important feature of transitional planar or nearly planar wall-bounded flows is indeed the separation of the full space into turbulent and laminar regions separated by sharp fluctuating interfaces, in the form of {\it turbulent spots\/} near the global stability threshold  $R_{\rm g}$ and more complicated laminar-turbulent {\it patterns\/} at larger values of the Reynolds number. 
Formally, this can be understood as resulting from a modulation in space of the strength of the SSP mechanism, active in turbulent domains and switched off in the surrounding laminar flow.
This modulation of the mechanism's intensity is in general detectable up to a limit above which a regime of uniform turbulence called {\it featureless\/}~\cite{Aetal86} is recovered.
The transition from modulated to featureless turbulence may then be marked by a well-defined threshold usually denoted $R_{\rm t}$, and laminar-turbulent coexistence be observed  in an interval $[R_{\rm g},R_{\rm t}]$ of finite width. 
For example, in plane Couette flow (PCF, the shear flow between counter-translating parallel plates at distance $2h$ and relative speed $2U$, for which $R:=Uh/\nu$), the transitional range extends from $R_{\rm g}\approx 325$ to $R_{\rm t}\approx 415$ and a periodic modulation of the turbulence intensity is observed in the form of bands alternatively laminar and turbulent, oblique with respect to the streamwise direction~\cite{Petal03,Ma16}.

Both around turbulent patches or between turbulent bands, near-laminar flow is the superposition of the base flow and large scale corrections believed to play an important role in the overall structure of turbulence~\cite{BT07}.
These corrections have components that do not average to zero in the wall-normal direction, which makes them capable of transporting coherent structures as a whole, subsequently acting on oblique spot growth~\cite{DS13,CM17}.
For that reason, they will be called {\it drift flows\/} in the following.
Oblique growth and the oblique laminar-turbulent organization clearly break the original spanwise symmetry of the problem statistically restored beyond $R_{\rm t}$.
The aim of this note is to bring hints on the origin of these flows and their relation to symmetry breaking at a local level, i.e. the Minimal Flow Unit (MFU), the scale introduced by Jim\'enez and Moin~\cite{JM91} below which coherent structures with sizable lifetimes are no longer observed.
The MFU is defined in the context of numerical simulations with wall-parallel periodic boundary conditions at distances $\ell_x$ and $\ell_z$ typically of the same order of magnitude as the wall-normal characteristic size $\ell_y$, viz. the gap $2h$ for PCF.
It is the privileged scale at which transitional coherent structures are studied~\cite{Getal09} and arguments from dynamical systems and chaos theory are developed~\cite{Ketal12}.

Great progress in the understanding of the SSP has been obtained thanks to Waleffe~\cite{Wa97} who built an analytical, relatively simple, model accounting for it within the Minimal Flow Unit (MFU) framework, called {\sc Wa97\/} in the following.
It was obtained as a first-harmonic truncation of a Galerkin expansion of the NSEs with stress-free boundary conditions in a plane Couette-like geometry, sometimes called {\it Waleffe flow\/}~\cite{Cetal16}.
A straightforward analysis using trigonometric basis functions yielded {\sc Wa97} as a system of 8 differential equations governing 8 mode amplitudes that was next reduced to a 4-dimensional system, the one studied in greatest detail.
The four amplitudes retained were explicitly associated to a mean flow distortion, the streaks and vortices amplitudes  and the amplitude of a combined mode effective in closing the system appropriately.
By construction {\sc Wa97}, whether reduced or not, preserves the spanwise symmetry of the flow configuration.
My purpose will be to generalize Waleffe's approach to allow for drift flows observed in numerical simulations~ \cite{DS13} or experiments~\cite{CM17}.
In some sense, this can be viewed as an analytical counterpart to the numerical approach developed by Kreilos {\it et al.\/}~\cite{Ketal14} who studied drifting patterns at MFU size in related transitional flows.

The Galerkin approach to be used is a weighted-residual method analyzing the problem at hand by expanding its fields and governing equations onto functional bases.
When pushed at high orders, it can serve as a numerical simulation method with good convergence properties~\cite{Fi72}.
It is however usually not developed as such in computational fluid dynamics and alternate methods are used, e.g.~\cite{Gi08}, more straightforward but rather working as black-boxes not amenable to analytical developments.
Here, the aim is therefore not to apprehend the {\it abstract\/} structure of state space  within the MFU framework in detail through the accurate determination of exact solutions to the NSEs like in \cite{Getal09}.
On the contrary, and much in the spirit of Waleffe's seminal work~\cite{Wa97}, I will attempt to uncover the {\it concrete\/} local source of mean-flow corrections involved in the symmetry breaking typically observed at transitional values of $R$.
To this aim, I will consider the Galerkin method rather as a systematic reductive modeling strategy of the primitive equations, achieved by truncating the expansion at the lowest possible but still significant order, so low that it can still be handled analytically while clear physical significance can be given to the mode amplitudes retained.

In accordance with the wide generality of the SSP for the base flows of interest, systems with similar structures can be derived with differences only appearing in the value of the coefficients.
Since trigonometric relations between the basis functions used to deal with the stress-free boundary conditions for Waleffe flow~\cite{Wa97} artificially kill some nonlinear interactions, in order to work with a slightly less restrictive case, I will consider standard PCF driven by  no-slip conditions at the plates.
On another hand, I will follow Waleffe in his restriction to a first-harmonic approximation of the MFU dynamics to describe the wall-parallel periodic dependence of the state variables.
Section~\ref{S2} gives a cursory presentation of the model, the full expression of which is given in the Supplement~\cite{Sup}.
Its main properties and virtues are then discussed in \S\ref{S3} before presenting some perspectives on laminar-turbulent patterning in transitional wall-bounded flows from a more general standpoint in \S\ref{S4}.

\section{The model\label{S2}}

The modeling approach starts from the velocity-vorticity formulation of the NSEs written for the perturbation to the base flow as detailed in~\cite{SH01}, p.~155ff.
Though the wall-normal and wall-parallel directions can be treated simultaneously in the Galerkin approach as originally done by Waleffe~\cite{Wa97}, here I will first deal with the wall-normal direction making use of results in~\cite{LM07,SM15}, and next with the wall-parallel direction and the periodic conditions corresponding to the MFU definition.
Simulations of PCF have shown that a representation of the flow at lowest significant order contains about 90\% of the perturbation energy for transitional Reynolds numbers~\cite[App.B]{Ma15}, accordingly I will just consider the corresponding minimal functional set, like for {\sc Wa97}:
\BE
\label{Exp1}\{u,w\} = \{U_0,W_0\}\>f_0(y) + \{U_1,W_1\}\> f_1(y),\quad\mbox{and}\quad v = V_1 \>g_1(y),
\EE
$\{u,w\}$ and $v$ being the wall-parallel and wall-normal perturbation velocity components,  respectively.
Polynomial bases introduced in~\cite{LM07,SM15} are particularly well adapted to the no-slip boundary conditions at $y=\pm1$, base flow $u_{\rm b} =y$, and low-order truncation~\cite{LM07}, namely%
\footnote{In the stress-free case, one has $f_0=1/\sqrt2$, $f_1= \sin \pi y/2$, $g_1=\cos \pi y/2$, and $u_{\rm b}\propto f_1$.} $f_0\propto (1-y^2)$, $ f_1\propto y \,(1-y ^2)$, and $g_1\propto (1-y^2)^2$.
Amplitudes $U_0$ and $W_0$ are attached to parabolic flow components that do not average to zero over $[-1,1]$ and clearly contribute to the drift flows mentioned in the introduction.

At this stage, amplitudes $\{U_{0,1},W_{0,1}\}$ and $V_1$, and the wall-normal vorticity components $Z_{0,1}=\partial_z U_{0,1} -\partial_x W_{0,1}$, are still functions of space~$(x,z)$ and time~$t$.
The partial differential equations expressing the NSEs, or rather their Orr--Sommerfeld part for $V_1$ and Squire part for $Z_{0,1}$ are given in the Supplement.
The next step is the MFU reduction, treated by a Fourier series expansion expressing the wall-parallel periodic boundary conditions at $\ell_x=2\pi/\alpha$ and $\ell_z=2\pi/\gamma$,
where $\alpha$ and $\gamma$ are the fundamental wavevectors of the wall-parallel MFU space dependence.
Here the expansion is truncated beyond  the first harmonic as in~\cite{Wa97} since we are not primarily interested in an accurate representation of the solution.
The generalization of Walleffe's ansatz reads: 
\BA
\NN \Psi_0 = - \overline U_0 z + \overline W_0 x + X_1 \sin\alpha x + X_2 \sin\gamma z +  X_1^u \cos\alpha x + X_1^w \cos\gamma z\\
\quad\mbox{}
+ X_3 \cos\alpha x\cos\gamma z + X_2^u \sin\alpha x\cos\gamma z
+ X_2^w \cos\alpha x \sin\gamma z + X_1^o \sin\alpha x \sin\gamma z,
\label{EPsi0}\\
 \NN \Psi_1= - \overline U_1 z + \overline W_1 x + X_4 \cos\alpha x + X_3^u \sin\alpha x + X_4^u \sin\gamma z + X_2^o \cos\gamma z\\
\quad\mbox{} + X_5 \sin\alpha x \cos\gamma z + X_5^u \cos\alpha x \cos\gamma z
+ X_3^w \sin\alpha x \sin\gamma z + X_3^o \cos\alpha x \sin\gamma z,
\label{EPsi1}\\
\NN \Phi_1 =  X_6 \cos\gamma z + X_4^w \sin\alpha x + X_5^w \sin\gamma z +  X_4^o \cos\alpha x\\
\quad\mbox{}
+ X_7 \cos\alpha x \sin\gamma z + X_6^u \sin\alpha x \sin\gamma z + X_6^w \cos\alpha x \cos\gamma z + X_5^o \sin\alpha x \cos\gamma z.
\label{EPhi1}
\EA
The velocity components are retrieved from the expression of the streamfunctions $\Psi_{0,1}$ and velocity potential $\Phi_1$ through:
\BE
U_0=-\partial_z \Psi_0,~~W_0=\partial_x \Psi_0,~~
U_1= - \partial_z \Psi_1-\beta \partial_x \Phi_1,~~ W_1 = \partial_x \Psi_1-\beta \partial_z \Phi_1,
\label{EUVW}
\EE
so that
$$
Z_{0,1}=-\Delta \Psi_{0,1}\qquad \mbox{and} \qquad V_1 = - \Delta \Phi_1,
$$
where $\Delta =\partial_{xx}+\partial_{zz}$ is the Laplacian in the plane of the flow and $\beta$ plays the role of a wall-normal wavevector (no-slip: $\beta = \sqrt3$, stress-free: $\beta=\pi/2$).
The terms $-\overline U_{0,1} z + \overline W_{0,1} x$  in (\ref{EPsi0},\ref{EPsi1}) correspond to the non-oscillatory mean-flow components, governed by appropriately averaged equations as discussed in~\cite{SH01}.
In {\sc Wa97}, only $\overline U_1$ and the set $\{X_1,\dots, X_7\}$ are present under different names  (equations~(8,9) in \cite{Wa97}), specifically: $M=1+\overline U_1$ (mean flow), $U=-\gamma X_2$ (streak amplitude), $V=-\gamma X_6$ (streamwise vortex amplitude), $A=\alpha X_1$,  $B=-X_3$, $C=-\alpha X_4$, $D=X_5$, and $E=-X_7$.
The justification for superscripts `$u$', `$w$', and `$o$', decorating the other sets of amplitudes amplitudes will appear in the next section.
Amplitudes $\overline U_0$ and $\overline W_0$ are the key ingredients in the extension of {\sc Wa97}. 

A set of 28 equations for the 28 unknowns is obtained by mere separation of harmonics.
It displays all the properties, lift-up, viscous dissipation, quadratic advection nonlinearities, linear stability of the base flow, expected from NSEs for wall-bounded shear flows within the MFU framework.
It formally reads:
\BE
\label{EG}
\mbox{$\frac{\rm d}{{\rm d} t}$} \mathcal Z + \mathcal L\, \mathcal Z = \mathcal N (\mathcal Z,\mathcal Z),
\EE
where the variable set $\mathcal Z$ can further be decomposed into:
\BE
\mathcal Z = \{\mathcal Y,\mathcal Y^u,\mathcal Y^w, \mathcal Y^o\} \equiv \big\{ \big\{\overline U_1,\mathcal X\big\},\big\{ \overline U_0, \mathcal X^u\big\},\big\{ \overline W_0, \mathcal X^w\big\},\big\{\overline W_1,X^o\big\} \big\},\label{EV}
\EE
where $\mathcal X=\{X_1,\dots X_7\}$, 
$\mathcal X^u=\{X_1^u,\dots X_6^u\}$, $\mathcal X^w=\{X_1^w,\dots X_6^w\}$, and 
$\mathcal X^o=\{X_1^o,\dots X_5^o\}$.
$\mathcal Y=\big\{\overline U_1,\mathcal X\big\}$ is precisely the set corresponding to {\sc Wa97}.
The full expression of System (\ref{EG}) is given in the Supplement where equations labelled~(n) are here named~(Sn).
An immediate inspection of this system shows that the subspace spanned by $\big\{\overline U_1,\mathcal X\big\}$ is closed, which means that, $\mathcal Z^{\mbox{\footnotesize\sc Wa97}} \equiv \big\{\mathcal Y, 0,0,0 \big\}$ is a consistent assumption solving the problem with equations for $\mathcal Y^u$,  $\mathcal Y^w$, and $\mathcal Y^o$ identically cancelling.
Here are four sample equations:
The first one (S27) governs the streamwise mean flow correction:
\BA
\NN\mbox{$\frac{\rm d}{{\rm d} t}$} {\overline U}_1 + \nu \bar p_1 {\overline U}_1 = \mbox{$\frac14$}\gamma \bar s_1\big[2\gamma^2( X_6 X_2 - X_1^w X_4^w ) \\
\label{EU1}\mbox{}+ \kappa^2 ( X_6^w X_2^w + X_5^o X_1^o - X_2^u X_6^u - X_3 X_7 )\big]\EA
where $\kappa^2=\alpha^2+\gamma^2$.
Once reduced reduced to {\sc Wa97}, it closely corresponds to Waleffe's equation (10-1) for $M=1+{\overline U}_1$~\cite{Wa97}, in the present notations:
\BE
\mbox{$\frac{\rm d}{{\rm d} t}$} {\overline U}_1 + \nu \bar p_1 {\overline U}_1 = \mbox{$\frac14$}\gamma \bar s_1\big[2\gamma^2 X_6 X_2 - \kappa^2 X_3 X_7\big].\label{EU1b}
\EE
My second sample is (S26) governing $\overline W_0$, a spanwise mean flow correction absent from {\sc Wa97}:
\BA
\NN\DDt \overline W_0 +\nu \bar p_0  \overline W_0 = \sfrac14 \alpha \bar s_0 [ 2\alpha^2( X_4  X_4^w - X_4^o X_3^u ) \\
\label{EW0}\mbox{}+ \kappa^2 (  X_5^o  X_5^u  + X_3^o  X_6^u  - X_5 X_6^w - X_7 X_3^w)].
\EA
The two last equations in this group, (S25) for $\overline U_0$ and (S28) for $\overline W_1$, follow the same simple pattern.
The third sample is (S2), the equation governing the {\it streak\/} amplitude $X_2$:
\BA
\NN
\mbox{$\frac{\rm d}{{\rm d} t}$} X_2 + \nu\kappa_0^\gamma X_2 = \bar b\, X_6 + s_0 \gamma \big( X_1^w {\overline W}_0 + \mbox{$\frac12$}\alpha (X_1 X_3 - X_1^u X_2^u)\big)\\
\NN\mbox{} + s_1 \big[ \gamma \big(X_2^o {\overline W}_1 + \beta X_6 {\overline U}_1\big) + \mbox{$\frac12$} \alpha\gamma \big((X_3^u X_5^u - X_4 X_5) + \beta^2 (X_4^o X_5^o-X_5^w X_6^w)\big)\\
\hspace{4em}\mbox{}- \mbox{$\frac12$}\alpha^2\beta (X_4X_7 + X_4^oX_3^o + X_3^u X_6^u + X_5^w X_3^w) \big)\big],\label{EX2}
\EA
with  $\kappa_0^\gamma = \gamma^2 + \bar p_0$.
The last sample is (S17) for the {\it streamwise vortex\/} amplitude $X_6$ ($\propto V$ in {\sc Wa97}) that generates the streaks by lift-up through the term $\bar b\, X_6$ in (\ref{EX2}):
\BA
\NN\mu^2_\gamma \DDt   X_6  + \nu\kappa'^4_\gamma  X_6   =    \alpha^2 r   (   X_1^u   X_5^u  + X_4   X_3 +  X_1   X_5 + X_3^u   X_2^u  )\\ 
\hspace{4em}\mbox{}+d_\gamma \left( X_5^w   X_2^w  -  X_4^o   X_1^o \right) +  e_\gamma( X_1^u   X_6^u  -  X_1   X_7 )- c_\gamma  X_6^w   \overline W_0 ,
\label{EX6}
\EA
with $\mu^2_\gamma = \gamma^2 + \beta^2$ and $\kappa'^4_\gamma=\gamma^4 + 2 \beta^2\gamma^2 + p_1$.
The values of constants appearing in the equations, $p_1$, $\bar p_{0,1}$, etc.  derive from the wall-normal part of the modeling of the considered flow configuration, thus here depending on whether no-slip or stress-free boundary conditions are used.
Once reduced to {\sc Wa97}, (\ref{EX2}) and (\ref{EX6}) read:
\BA
\NN\DDt X_2 + \nu\kappa_0^\gamma X_2 = \bar b\, X_6 + \mbox{$\frac12$} \gamma \alpha s_0 X_1 X_3 + s_1 \big( \beta\gamma X_6 {\overline U}_1\\
 - \mbox{$\frac12$} \alpha\gamma X_4 X_5-\mbox{$\frac12$}\alpha^2\beta X_4X_7\big)\label{EX2b}
\EA
and
\BE
\mu^2_\gamma \DDt X_6 + \nu\kappa'^4_\gamma  X_6 = \alpha^2 r ( X_4 X_3 + X_1 X_5) - e_\gamma X_1 X_7 \label{EX6b}
\EE
strictly corresponding to Waleffe's~(10-2) and (10-3).
When comparing his system to the corresponding one extracted from (S1)--(S28), a single difference appears in equation (S9) for $X_4$ that reads
\BA
\NN\DDt X_4 + \nu \kappa^2_\alpha  X_4 \!= -\alpha  b X_1\! + s_1(\alpha\gamma X_2 X_5 - 2 \alpha X_1\overline U_1 -\beta\gamma^2 X_2 X_7)\\
\mbox{} -  \sfrac12 s_4 \gamma^2 X_6   X_3. \label{EX4}
\EA
In {\sc Wa97}, $X_4$ is variable $C$ and the corresponding equation is (10-6) with the same terms as in (\ref{EX4}) but lacks the last one, $X_3X_6$, i.e. $BV$, that disappears owing to an accidental cancellation from trigonometric relations as noticed earlier.
The detailed consequences of this observation have however not been scrutinized in detail.

Before considering the virtues and limitations of model (\ref{EG}) in the following sections, let me stress that, within the 1st-harmonic MFU assumption, its expression and detailed structure are quite general so that its applicability is not restricted to PCF or Waleffe flow.
On the contrary, it should rather be understood as implementing the SSP on an extended footing that includes drift flows.

\section{Translational invariance and the generation of drift flows\label{S3}}
The 28 variables in $\mathcal Z$ is the most general ensemble compatible with the first-harmonic approximation.
As detailed in the Supplement, equations in (\ref{EG}) are individually rather complicated but with clear physical meanings.
Terms with $\nu$ in factor of specific expressions of the Laplacian obviously account for viscous dissipation and lift-up, already identified as $\bar b X_6$ on the right hand side of (\ref{EX2}), acts similarly on other mode pairs explicitly periodic in $z$, e.g. $X_7$ as a source term for $X_3$.
Conservation of the kinetic energy by quadratic terms is expected from the way the model is derived.
It is indeed fulfilled but the check remains technically cumbersome. 

More importantly, the choice $\mathcal Z^{\mbox{\footnotesize \sc Wa97}} = \{\mathcal Y, 0,0,0 \big\}$ is associated with specific spatial resonances between the different flow components.
This resonance condition can be retrieved in each and every solution to the full system,  whatever its time dependence, by performing an arbitrary time-independent translation $x\mapsto x+x_0$, $z\mapsto z+z_0$.
Similar observations have been made in the literature, see~\cite{Ketal14} and references quoted.
Here I will take a down-to-earth but instructive viewpoint and first note that this implies relations between the equations of the full system.
For example, a translation by $\ell_x/4$ amounts to performing the changes $\cos\alpha x\mapsto\sin\alpha x$ and $\sin \alpha x\mapsto-\cos \alpha x$, which straightforwardly explains the similarity of equations for $X_1$ and $X_1^u$, (S1) and (S3), $X_3$ and $X_2^u$, (S5) and (S6), etc. with identical coefficients and signs modifications linked to the minus sign in the second change.
An immediate consequence is that the dynamics restricted to $\mathcal Z^{\mbox{\footnotesize\sc Wal97},u}=\{ \mathcal Y, \mathcal Y^u, 0,0\}$ is also closed.
The case of $z$-translation can be treated in the same way, showing that the subspace $\mathcal Z^{\mbox{\footnotesize\sc Wal97},w}=\{ \mathcal Y, 0, \mathcal Y^w,0 \}$ is similarly invariant.
Inspecting the full expression of (S1--S28) finally shows that the subspace $\mathcal Z^{\mbox{\footnotesize\sc Wa97},o} = \{ \mathcal Y, 0,0, \mathcal Y^o\}$ is also invariant but with no obvious relation to translational properties.
 
On general grounds, the knowledge of the structure of phase space takes advantage of the stability properties of solutions known.
As a consequence of the identification of invariant subspaces above and the quadratic nature of the nonlinearities, it follows from standard linear analysis that the stability operator around a solution in $\mathcal Z^{\mbox{\footnotesize\sc Wa97}}$ has a block diagonal structure.
The first $8\times8$ block accounts for the stability of the solution as if the system was restricted to $\big\{\overline U_1,\mathcal X\big \}$, as dealt with in~\cite{Wa97}.
It corresponds to {\it amplitude\/} perturbations.
The two next $7\times7$ blocks are for infinitesimal perturbations living in $\big\{ \overline U_0, \mathcal X^u\}$ and $\big\{ \overline W_0, \mathcal X^w\}$.
These subspaces being associated with translations as discussed above, the related linear modes correspond to {\it phase\/} perturbations.

For example, let us consider the effect of an infinitesimal translation $z\mapsto z+\delta_z$ on an arbritrary  solution $\mathcal Z =\{\mathcal Y,0,0,0\}$.
At leading order the solution reads $\mathcal Z+\delta \mathcal Z$ with
$\delta Z = \{0,0,\epsilon_z \mathcal Y'^w, 0\}$ with $\epsilon_z=\gamma\delta_z$.
The components of $\mathcal Y'^w$ are $\overline W_0=0$,  $ X_1'^w= X_2$, $ X_2'^w=-X_3 $, $X_3'^w= -X_5$, $X_4'^w = 0$, $ X_5'^w =-X_6 $, and $ X_6'^w = X_7$.
It is indeed readily checked that the right hand side of equation equation (S18) governing $X_4^w$ cancels identically for such a perturbation, which implies $X_4^w\to 0$.
Next for $X_4^w=0$, it is verified that the right hand side of~(\ref{EW0}) also vanishes, so that $\overline W_0\to0$.
In addition, the equations governing all the non-zero components of $\mathcal Y'^w$ are identical to the equations for the corresponding component of $\mathcal Y$ apart from appropriate sign changes and, finally, insertion of perturbation $\epsilon_z\mathcal Y'^w$ in equation (\ref{EU1}) for $\overline  U_1$ yields
\BE
\mbox{$\frac{\rm d}{{\rm d} t}$} {\overline U}_1 + \nu \bar p_1 {\overline U}_1 = \mbox{$\frac14$}\gamma \bar s_1\big[2\gamma^2 X_6 X_2
- \kappa^2 X_3 X_7 (1+\epsilon_z^2)\big],
\label{EU1c}\EE
which shows that the dynamics of $\overline U_1$ is preserved at leading order.
Accordingly, perturbations corresponding to an infinitesimal $z$-translation are neutral  and do not generate drift flow $\overline W_0$, as expected. 
The same argument can be developed for infinitesimal streamwise translations, with identification of the corresponding perturbation mode $\mathcal Y'^u$ and proof of the absence of related $\overline U_0$.
However, perturbations along the so-obtained eigenvectors $\mathcal Y'^w$ and $\mathcal Y'^u$, while neutral, are extremely special and it is immediately seen that arbitrary perturbations are expected to generate some non trivial drift flow $\big( \overline U_0,\overline W_0\big)$.
It suffices to look at (\ref{EW0}), for convenience rewritten by dropping all irrelevant higher order terms as:
\BE
\DDt \overline W_0 +\nu \bar p_0  \overline W_0 = \sfrac14 \alpha \bar s_0 [ 2\alpha^2  X_4  X_4^w - \kappa^2 (X_5 X_6^w + X_7 X_3^w)]\label{EW0b}
\EE
to see how perturbations within subspace $\big\{\overline W_0,\mathcal X^w\big\}$ but with $\mathcal X^w \not\equiv \epsilon_z \mathcal X'^w$, i.e. $X_4^w=\epsilon_1\ne0$, $X_6^w=\gamma\delta_z X_7 +\epsilon_2$, $X_5^w=-\gamma\delta_z X_6 +\epsilon_3$, introduce sources terms for $\overline W_0$, generating a response of the same order of magnitude that comes and feeds back into the whole system.
Going back to the definitions of the different variables, assuming $\epsilon_{1,2,3} \ne0$ means that arbitrary infinitesimal perturbations $X_4^w$, $X_6^w$, and $X_3^w$ can resonate with already present flow components, $X_4$, $X_5$, and $X_7$ to produce some drift flow $\overline W_0$ as soon as they do not strictly derive from an infinitesimal spanwise translation.
Let us just consider the contribution of $X_4 X_4^w$ to the r.h.s. of (\ref{EW0b}) since $X_4^w\equiv0$ for an infinitesimal translation:
Returning to (\ref{EPsi1}), we see that $X_4$ is the amplitude of a spanwise velocity component $-\alpha X_4\sin\alpha x$ of the nontrivial state of interest, hence of order one, that interacts resonantly by lift-up with an infinitesimal wall-normal velocity component $(-\alpha^2/\beta) X_4^w$ as obtained from (\ref{EPhi1}) to produce a distortion $\overline W_0$. The others contributions $X_6^w$ and $X_3^w$ departing from strict infinitesimal translation would be analyzed in the same way with production of some $\overline W_0$ as a net result.
In turn the so-produced $\overline W_0$ of order $\epsilon$ feeds back into the dynamics of the set $\mathcal X^w$ also of order $\epsilon$, while corrections to dynamics of set $\mathcal X$ are of order $\epsilon^2$, as a consequence of the block structure of the linear stability operator.
Unfortunately, without specifying the nontrivial state of interest---which is clearly beyond the scope of this study---it is not possible to go further and decide whether system $\mathcal Y^w=\big\{\overline W_0, \mathcal X^w\big\}$ is stable or unstable, i.e. whether it has exponentially growing solutions in addition to the neutral phase mode that exists in all circumstances.
The same reasoning would also separately apply to streamwise perturbations.
What precedes leads us to suspect that the study of solutions obtained in a MFU context, either in a model like (S1--S28) or in the full NSEs, lacks an important ingredient if symmetries are imposed that forbid the existence of drift flows.

\section{Discussion and perspectives\label{S4}}

Periodic boundary conditions inherent in the MFU assumption maintain the fiction of a solution that would be uniformly developed in space.
In actual systems with wall-parallel dimensions much larger than the wall-normal scale, itself typical of the size of the MFU, the intensity of the SSP can be modulated, especially in the turbulent spot regime around~$R_{\rm g}$  where the turbulence level varies from 0 to 1 in space, and in oblique banded laminar-turbulent patterns up to~$R_{\rm t}$.
In the stability analysis sketched above, spanwise and streamwise translations could be treated separately.
This is no longer the case more generally since the corresponding drift flows are coupled by the continuity equation, an important condition at the heart of the Duguet--Schlatter argument about the obliqueness of laminar-turbulent interfaces~\cite{DS13}, here expressed as~\cite{LM07}:
$$
\partial_x \overline U_0 + \partial_z \overline W_0 = 0,
$$
and automatically fulfilled thanks to (\ref{EUVW}).
Modulations to the SSP intensity have to be understood as perturbations brought to a {\sc Wa97} solution that, as is readily verified, must include all the components of its extended representation in the present model.
For convenience (\ref{EG}) can be rewritten by separating the mean flows $\overline{\mathcal U} = \big\{\overline U_0, \overline W_0, \overline U_1, \overline W_1\big\}$ from the rest of the amplitudes $\underline{\mathcal X} = \big\{\mathcal X,\mathcal X^u,\mathcal X^w, \mathcal X^o \big\}$:
\BA
\label{EG1}
\mbox{$\frac{\rm d}{{\rm d} t}$} \underline{\mathcal X} + \mathcal L\, \underline{\mathcal X} + \mathcal M\big(\,\overline{\mathcal U}\,\big)\underline{\mathcal X}=\mathcal N \big(\underline{\mathcal X},\underline{\mathcal X}\big),\\
\label{EG2}
\mbox{$\frac{\rm d}{{\rm d} t}$}\overline{\mathcal U} + \mathcal L' \overline{\mathcal U} = \mathcal N' \big(\underline{\mathcal X},\underline{\mathcal X}\big),
\EA
highlighting the origin and role of $\overline{\mathcal U}$.

Assuming that this system only describes the small-scale (MFU) flow in a simplified way and admitting further that this local solution can experience modulations that perturb the fine tuning of SSP modes, we see that drift flows inevitably appear as nontrivial responses to Reynolds stresses on the r.h.s. of (\ref{EG2}) induced by resonance mismatches pointed out in \S\ref{S3} above. 
These drift flows then feed back into the rest of the solution {\it via\/} the term $\mathcal M \left(\overline{\mathcal U}\right) \underline{\mathcal X}$ in (\ref{EG1}); typical examples are the terms with a factor $\overline W_0$ in the equations governing the amplitudes of the  two most crucial ingredients to the SSP, (\ref{EX2}) for the streaks $X_2$ and (\ref{EX6}), for the streamwise vortices $X_6$.

When dealing with modulations, we have to face the difficulty that, owing to the sub-critical character of transition, there is no systematic multiscale approach available.
Nontrivial solutions emerge abruptly and steep interfaces in physical space separate different flow regimes, in sharp contrast with the case of supercritical bifurcations as pointed out by Pomeau \cite[\S4]{Po15}.
The first reason is that there is no linear marginal stability condition to work with:
the coherence length that controls the diffusion of modulations near threshold is indeed directly obtained from its curvature at the critical point.
The second reason is that supercriticality implies a controllable saturation of the solution's amplitude.
Both circumstances permit a rigorous and systematic perturbation approach~\cite[Chaps.~8--10]{Ma90}, and none holds in the present case.

The spatial coexistence of laminar and turbulent flows is particularly difficult to apprehend from the primitive equations.
An approach {\it via\/} analogical modeling in terms of reaction-diffusion (RD) systems~\cite{Mu93},  developed by Barkley~\cite{Ba11,Ba16}, has been particularly fruitful to account for the transitional range of pipe flow.
In that model, the production of turbulence was considered as the result of a reaction and diffusion was introduced phenomenologically to treat the spatial coexistence of the two states, laminar or turbulent.
Soon after the earliest developments of that work, I used the same RD framework but in the context of a Turing instability, i.e. a pattern-forming mechanism controlled by diffusion rates with sufficiently different orders of magnitude~\cite{Mu93}.
In my model~\cite{Ma12}, the local reaction terms were expressed using the reduced (4-dimensional) {\sc Wa97} model~\cite{Wa97}, while its variables were allowed to diffuse with widely different turbulent viscosities in one direction of space.
As a result, a Turing bifurcation was obtained at decreasing $R$, defining a threshold $R_{\rm t}$ below which a pattern was present down to some $R_{\rm g}$ corresponding to a general breakdown toward laminar flow.
Whereas it seems reasonable to use the variables in $\underline{\mathcal X}$ to treat turbulence production at a local scale, the structure of (\ref{EG1},\ref{EG2}) clearly shows that the simple heuristic assumption of a diffusion {\it via\/} turbulent viscosities is unable to properly render the possible role of drift flows on pattern formation.
On the other hand, a model equivalent to the spatiotemporal Galerkin system described in \S1 of the Supplement  was numerically studied by Lagha and myself in~\cite{LM07b}.
Filtering out the small scales, we could next determine the dynamics of large-scale flows,  in particular their drift-flow component $\big(\overline U_0,\overline W_0\big)$ around turbulent spots.
They were obtained analytically as a response to Reynolds stresses given from the outside, not as stemming from some local dynamics possibly obtained within a MFU framework as examined here.

We can therefore infer that a combination of the two approaches, small scale dynamics including the feedback of large scales flows, should provide a satisfactory, now self-contained, description of laminar-turbulent coexistence in the transitional regime.
Numerical simulations of Galerkin models truncated at different levels however suggest that  the lowest nontrivial, three-field, level is insufficient to recover an organized laminar-turbulent  band pattern, for PCF~\cite{SM15} as well as for Waleffe flow~\cite{Cetal16}, and that we are requested to consider at least seven fields in order to obtain oblique bands in a $[R_{\rm g},R_{\rm t}]$ range of finite width~\cite{SM15}.
Working with a higher level model at the MFU scale, further incorporating the effect of space modulations, and of course simplifying the cumbersome so-obtained model appropriately, e.g. through adiabatic elimination of fast variables, is likely the only way to really {\it explain\/} the occurrence of laminar-turbulent pattern analytically.
A RD picture~\cite{Ba11,Ba16,Ma12} would emerge, mostly directed at the understanding of the  transition from featureless turbulence to banded turbulence at $R_{\rm t}$ upon decreasing $R$.
It would be derived from the NSEs and would replace the naive introduction of turbulent viscosities by a clean account of drift flows, hopefully containing the mechanism for a Turing instability.   
The approach is not limited to PCF or Waleffe flow and should provide a generic interpretation to laminar-turbulent coexistence in the transitional range for a wide range of wall-bounded flows of practical interest, as can be anticipated from the universal structure of Galerkin approximations to the NSEs, the relevance of the SSP in producing nontrivial states already at the MFU scale and moderate Reynolds numbers, and the ubiquitous presence of drift flows.
\bigskip

\noindent Acknowledgments: I would like to thank Profs. G.~Kawahara and M.~Shimizu (Osaka University, Japan) and  Dr. Y.~Duguet (LIMSI, France) for discussions related to this work within the framework of the {\sc TransTurb}  JSPS--CNRS exchange program.

\bibliography{bibliography}

\end{document}


\title{Supplement}
\date{}
\maketitle

\section{Galerkin three-field model}

As explained in the text, we model PCF by severely truncating a Galerkin expansion of the no-slip problem starting with the velocity-vorticity formulation of Navier--Stokes equations written for the perturbation to a general streamwise laminar base flow $\mathbf u_{\rm b}=u_{\rm b} \mathbf{\hat x}$~\cite{SH01}. 
These equations read:
\BAN
(\partial_t + u_{\rm b} \partial_x) \nabla^2 v -u''_{\rm b}\partial_x v - \nu \nabla^4 v &=& \mathcal N_v\,, \label{eq_vv}\\
(\partial_t + u_{\rm b} \partial_x) \zeta + u'_{\rm b}\partial_z v -\nu \nabla^2\zeta &=& \mathcal N_\zeta\,. \label{eq_zeta}
\EAN
where $v$ is the wall-normal velocity component, $\zeta=\partial_z u-\partial_x w$ is the wall-normal vorticity component, $u$ ($w$) being the streamwise (spanwise) wall-parallel perturbation velocity component.
The equations are written for a general base flow $u_{\rm b}$, the primes indicating differentiation w.r. to direction~$y$.
For PCF, $u_{\rm b}=y$, hence $u'_{\rm b}=1$ and $u''_{\rm b}=0$.
The kinematic viscosity $\nu$ is numerically equal to $1/R$ when $h$ and $h/U$ are used as length and time units to write the equations in dimensionless form.
$\mathcal N_v$ and $\mathcal N_\zeta$ are complicated, formally quadratic expressions:
\BAN
\mathcal N_v&=& \partial_{xy}\mathcal{S}_x+\partial_{yz} \mathcal{S}_z - (\partial_{xx} + \partial_{zz}) \mathcal{S}_y\,,\\
\mathcal N_\zeta&=&-\partial_z \mathcal{S}_x+\partial_x \mathcal{S}_z\,,
\EAN
with
$$
\mathcal{S}_x=u \partial_x u + v \partial_y u + w \partial_z u,\qquad
\mathcal{S}_y=u \partial_x v + v \partial_y v + w \partial_z v,\qquad
\mathcal{S}_z=u \partial_x w + v \partial_y w + w \partial_z w.
$$

The perturbation velocity field is then expanded onto a convenient orthonormal polynomial basis as explained in~\cite{SM15} from which extract the lowest-order consistent set $\{ V_1, Z_0, Z_1\}$: 
\BAN
&&\NN\left(\Delta -\beta^2\right)\partial_t V_1 - \nu\left( \Delta^2 - 2 \beta^2 \Delta + p_1\right) V_1 =\mbox{} - \big(q \Delta -\bar r \,\big) \big( \partial_x (U_0 V_1 ) +\partial_z ( W_0 V_1 )\big) \\
&&\qquad\mbox{}+ r \left[\partial_{xx} \left( U_1 U_0 \right) + \partial_{xz}\left( U_1 W_0 + U_0 W_1 \right)+\partial_{zz} \left( W_0 W_1 \right)\right],\\
&&\NN\partial_t  Z_0 
 +b\, \partial_x  Z_1 + \bar b\, \partial_z V_1 - \nu \left( \Delta - \bar p_0\right)  Z_0 = \makebox{} s_0\left(\partial_{xz} \left( U_0 ^2 - W_0 ^2 \right) + (\partial_{zz}-\partial_{xx})( U_0 W_0 )\right] \\
&&\NN\qquad\mbox{}+s_1\left[\partial_{xz} \left( U_1 ^2 - W_1 ^2 \right) + (\partial_{zz}-\partial_{xx})( U_1 W_1 )\right] -\bar s_0 \left[\partial_z( U_1 V_1 ) -\partial_x( W_1 V_1 ))\right],\\
&&\NN\partial_t  Z_1 +b\, \partial _x  Z_0 - \nu \left( \Delta - \bar p_1\right)  Z_1 =\mbox{} 2s_1\left[\partial_{xz} ( U_1  U_0 - W_1 W_0 ) + (\partial_{zz}-\partial_{xx})( U_1 W_0 + U_0 W_1 )\right] \\
&&\NN\qquad\mbox{}-\bar s_1 \left[\partial_z( U_0 V_1 ) -\partial_x( W_0 V_1 )\right],
\EAN
where $\Delta =\partial_{xx}+\partial_{zz}$ is the Laplacian in the plane of the flow.
The stress-free and no-slip versions of this system have the same overall structure, only differing in the numerical values of the coefficients~\cite{LM07,SM15}. For linear terms:
$$
\begin{tabular}{ccccccc}
\vphantom{$\sqrt{\frac17}$} & $\beta$ & $\bar p_0$ & $\bar p_1$ & $p_1$ & $b$ & $\bar b$\\
no-slip \vphantom{${\sqrt{\frac17}}^2_2$}  & $\sqrt3\approx 1.73$ & $\frac52$ & $\frac{21}2$ & $\frac{63}2$ & $\frac1{\sqrt7} \approx 0.38$ & $\frac{3\sqrt3}{2\sqrt7}\approx 0.98$\\[1.5ex]
stress-free\vphantom{${\sqrt{\frac17}}^2_2$} & $\frac\pi2\approx 1.57$&$0$ &$(\frac\pi2)^2\approx2.47$& $(\frac\pi2)^4\approx6.1$ & $\frac1{\sqrt2} \approx 0.71,$ & $\frac{\pi}{2\sqrt2}\approx 1.11$\\[1.5ex]
 \end{tabular}
$$
from which it can be seen that the velocity profile of the stress-free case is much less dissipative  than the no-slip profile as expected owing to the absence of boundary layers close to the walls.\\
For nonlinear terms:
$$
\begin{tabular}{ccccc}
& $q$ & $ r $ & $\bar r$\\[1.5ex]
no-slip\vphantom{${\sqrt{\frac17}}^2_2$} & $\frac{5\sqrt{15}}{22}\approx0.88$ & $\frac{\sqrt{5}}2\approx1.12$& $-\frac{\sqrt{135}}{4}\approx -2.9$\\[1.5ex]
stress-free\vphantom{${\sqrt{\frac17}}^2_2$}  & $\frac1{\sqrt2}\approx0.71$ & $\frac\pi{2\sqrt2}\approx1.11$ & $-\frac{\pi^2}{4\sqrt2}\approx-1.74$\\[1.5ex]
 \end{tabular}
 $$
 and
$$
\begin{tabular}{ccccc}
& $s_0$ & $s_1$ & $\bar s_0 $ & $\bar s_1$ \\[1.5ex]
no-slip\vphantom{${\sqrt{\frac17}}^2_2$} & $\frac{3\sqrt{15}}{14}\approx0.83$ & $\frac{\sqrt{15}}{6}\approx0.65$ & $ \frac{\sqrt{5}}{4}\approx0.56$& $-\frac{3\sqrt{5}}{4}\approx-1.68$\\[1.5ex]
stress-free \vphantom{${\sqrt{\frac17}}^2_2$}& $\frac12$ &  $\frac12$ & $0$ & $-\frac{\pi}{2\sqrt2}\approx -1.11$ \\[1.5ex]
 \end{tabular}
 $$
Minor differences with coefficients given in \cite{LM07} may be noticed, all stemming from the fact that we use the velocity-vorticity formulation of \cite{SM15} rather than the Navier--Stokes equations in primitive variables and a subsequent treatment of the pressure field in \cite{LM07}. 

\section{First-harmonic approximation in the MFU context}
The model involves wave-vectors $\alpha=2\pi/\ell_x$ and $\gamma=2\pi/\ell_z$ as parameters, $\ell_x$ and $\ell_z$ being the wall-parallel dimensions of the MFU.
The first-harmonic basic guess is repeated here for convenience:
\BAN
\NN && \Psi_0 =\mbox{} - \overline U_0 z + \overline W_0 x + X_1 \sin\alpha x + X_2 \sin\gamma z +  X_1^u \cos\alpha x + X_1^w \cos\gamma z\\
\NN &&\quad\mbox{}+ X_3 \cos\alpha x\cos\gamma z + X_2^u \sin\alpha x\cos\gamma z + X_2^w \cos\alpha x \sin\gamma z+ X_1^o \sin\alpha x \sin\gamma z,\\
&& \NN \Psi_1=\mbox{} - \overline U_1 z + \overline W_1 x + X_4 \cos\alpha x + X_3^u \sin\alpha x + X_4^u \sin\gamma z + X_2^o \cos\gamma z\\
\NN &&\quad\mbox{} + X_5 \sin\alpha x \cos\gamma z + X_5^u \cos\alpha x \cos\gamma z
 + X_3^w \sin\alpha x \sin\gamma z + X_3^o \cos\alpha x \sin\gamma z,\\
\NN &&\Phi_1 = X_6 \cos\gamma z+  X_4^w \sin\alpha x + X_5^w \sin\gamma z  + X_4^o \cos\alpha x\\
\NN &&\quad\mbox{}+ X_7 \cos\alpha x \sin\gamma z + X_6^u \sin\alpha x \sin\gamma z + X_6^w \cos\alpha x \cos\gamma z  + X_5^o \sin\alpha x \cos\gamma z.
\EAN
The velocity and wall-normal vorticity components deriving from these fields read:
\BAN
U_0 = \overline U_0 - \partial_z \widetilde \Psi_0,&& W_0 = \overline W_0 + \partial_x \widetilde \Psi_0,\\
U_1 = \overline U_1 - \partial_z \widetilde \Psi_1 -\beta  \partial_x \Phi_1, &\qquad V_1=-\Delta\Phi_1,\qquad& W_1 = \overline W_1 + \partial_x \widetilde \Psi_1 -\beta \partial_z \Phi_1,\\
Z_0=\partial_z U_0 -\partial_x W_0 = -\Delta \widetilde  \Psi_0,&&
\,\, Z_1=\partial_z U_1 -\partial_x W_1 = -\Delta \widetilde \Psi_1\,,
\EAN
where $\widetilde \Psi_0$ and $\widetilde \Psi_1$ denote the periodically varying parts of $\Psi_0$ and $\Psi_1$.
In order to simplify the expressions of some coefficients, we introduce:
 $$
\begin{tabular}{ l l l l l l}
 $\kappa^2=\alpha^2+\gamma^2$, &\quad $\bar\alpha = \alpha/\kappa$, &\quad$\bar\gamma= \gamma/\kappa$, &\quad $\tau=\bar \gamma/\bar \alpha$,&\quad $g=\bar\alpha^2-\bar\gamma^2$, &\quad $g'=2\bar\alpha\bar\gamma$,
 \end{tabular}
 $$
where $\tau=\tan\theta$, $g=\cos 2\theta$, and $g'=\sin2\theta$, relate to the aspect ratio of the MFU.\\
Equations have been derived using {\sc Mathematica}.

\subsection{Equations stemming from the dynamics of $\Psi_0$}
\BA
\NN \DDt X_1 + \nu\bar\kappa^2_\alpha X_1 & = &\alpha b X_4 + s_0 \left (\alpha X_1^u \overline U_0 + \sfrac12\alpha\gamma\left( X_1^w X_2^w - X_2 X_3 \right) \right)\\
\NN &&\mbox{}+ s_1 \left( \alpha X_4 \overline U_1 +\sfrac12\alpha\gamma\left( X_2^o X_3^o - X_4^u X_5^u \right) -\sfrac12\beta\gamma^2\left( X_2^o X_5^o + X_4^u X_6^u \right)\right)\\
&&\mbox{}- s_2\left( \alpha  X_4^o  \overline W_1 +\sfrac12\alpha\beta\gamma\left( X_6 X_7 - X_5^w X_6^w \right)+\sfrac12\gamma^2\left(X_6 X_5 + X_5^w X_3^w \right)\right),
\label{E01}\\
\NN \DDt X_2 + \nu\bar\kappa^2_\gamma X_2 & = & \bar b X_6 + s_0 \left(\gamma X_1^w \overline W_0 +\sfrac12\alpha\gamma\left( X_1 X_3 - X_1^u X_2^u \right)\right)\\
\NN &&\mbox{} + s_1 \left(\gamma X_2^o \overline W_1 +\beta\gamma X_6 \overline U_1 + \sfrac12\alpha\gamma\left( X_3^u X_5^u - X_4 X_5\right)+\sfrac12\alpha\beta^2\gamma \left( X_4^o  X_5^o - X_4^w X_6^w \right)\right.\\
 &&\left.\hspace{8ex} \mbox{}- \sfrac12\alpha^2\beta \left( X_4 X_7 + X_4^o X_3^o + X_3^u X_6^u + X_4^w X_3^w \right)\right),
\label{E02}\\
\NN \DDt X_1^u + \nu \bar\kappa^2_\alpha X_1^u & = & - \alpha b X_3^u - s_0 \left( \alpha X_1 \overline U_0 +\sfrac12 \alpha \gamma \left( X_1^w X_1^o - X_2 X_2^u \right) \right)\\
\NN &&\mbox{}- s_1 \left(\alpha X_3^u \overline U_1 +\sfrac12\alpha\gamma\left ( X_2^o X_3^w - X_4^u X_5 \right) + \sfrac12\beta\gamma^2 \left( X_2^o X_6^w + X_4^u X_7 \right) \right)\\
&&\mbox{}+ s_2\left (\alpha X_4^w \overline W_1 +\sfrac12\alpha\beta\gamma\left ( X_6 X_6^u - X_5^w X_5^o \right) - \sfrac12\gamma^2\left( X_6 X_5^u + X_5^w X_3^o \right) \right),
\label{E03}\\
\NN \DDt X_1^w + \nu\bar\kappa^2_\gamma X_1^w & = & - \bar b X_5^w - s_0 \left( \gamma X_2 \overline W_0 + \sfrac12\alpha\gamma\left( X_1 X_2^w - X_1^u X_1^o \right)\right)\\
\NN &&\mbox{}- s_1 \left(\gamma X_4^u \overline W_1 +\beta\gamma X_5^w \overline U_1 + \sfrac12\alpha\gamma\left( X_3^u X_3^o - X_4 X_3^w 
\right) + \sfrac12\alpha\beta^2\gamma \left( X_4^o  X_6^u - X_4^w X_7 \right)\right.\\
&&\hspace{8ex}\left. \mbox{}+\sfrac12\alpha^2\beta \left( X_4 X_6^w +  X_4^o  X_5^u + X_3^u X_5^o + X_4^w X_5 \right)\right),
\label{E04}\\
\NN\DDt  X_3  + \nu \bar\kappa^2_{\alpha\gamma}  X_3  &  =    & -\alpha b  X_5  -\gamma  \bar b   X_7 - s_0  \left(\alpha X_2^u   \overline U_0  + \gamma  X_2^w   \overline W_0 -\alpha\gamma g X_1   X_2 \right)\\
\NN &&\mbox{}
-  s_1  \left(\alpha X_5   \overline U_1  + \gamma  X_3^o  \overline W_1 + \beta\gamma  X_7   \overline U_1 
-\alpha\gamma g  X_3^u  X_4^u +\sfrac12{g'}^2 \kappa^2 \beta  X_2^o   X_4^o \right)\\
&&\mbox{}
 +\alpha  s_2  X_5^o  \overline W_1 -  \sfrac14{g'}^2 \kappa^2  s_3   X_4   X_6 
 -\alpha\beta\gamma(\bar\alpha^2 \bar s_0 + g\beta  s_1 )  X_5^w   X_4^w ,
\label{E05}
\EA
\BA
\NN \DDt X_2^u + \nu\bar\kappa^2_{\alpha\gamma}  X_2^u  &  =    & \alpha b  X_5^u      -  \gamma     \bar b   X_6^u +  s_0  \left(\alpha X_3   \overline U_0  - \gamma  X_1^o   \overline W_0 -\alpha\gamma g  X_1^u   X_2 \right)\\
\NN&&\mbox{}+  s_1  \left(\alpha  X_5^u   \overline U_1  - \gamma  X_3^w  \overline W_1 -\beta\gamma  X_6^u   \overline U_1 -\alpha\gamma g  X_4   X_4^u  - \sfrac12{g'}^2 \kappa^2\beta  X_4^w   X_2^o \right)\\
 &&\mbox{} -\alpha  s_2  X_6^w  \overline W_1  - \sfrac14{g'}^2 \kappa^2 s_3   X_3^u   X_6  + \alpha\beta\gamma(\bar\alpha^2 \bar s_0 + g \beta  s_1 )  X_4^o   X_5^w, 
\label{E06}\\
\NN\DDt  X_2^w  + \nu\bar\kappa^2_{\alpha\gamma}  X_2^w  &  =    &  \gamma  \bar b   X_6^w  - \alpha b  X_3^w + s_0 \left(\gamma  X_3   \overline W_0  - \alpha  X_1^o   \overline U_0 - \alpha\gamma g  X_1   X_1^w \right)\\
\NN &&\mbox{}
+  s_1  \left(\gamma  X_5^u   \overline W_1 -\alpha  X_3^w   \overline U_1 + \beta \gamma  X_6^w   \overline U_1   - \alpha\gamma g  X_3^u   X_2^o   - \sfrac12{g'}^2\kappa^2 \beta   X_4^u   X_4^o \right) \\
&&\mbox{}  +\alpha  s_2  X_6^u  \overline W_1 -  \sfrac14{g'}^2 \kappa^2  s_3  X_4   X_5^w +\alpha\beta\gamma(\bar\alpha^2 \bar s_0 + g\beta  s_1 )  X_4^w   X_6 ,
\label{E07}\\
\NN \DDt  X_1^o  + \bar\kappa^2_{\alpha\gamma}  X_1^o  &  =    & \alpha b  X_3^o  + \gamma  \bar b   X_5^o + s_0  \left(\gamma  X_2^u   \overline W_0 + \alpha  X_2^w   \overline U_0  +\alpha\gamma g  X_1^u   X_1^w \right)\\
\NN &&\mbox{}+  s_1  \left(\gamma  X_5  \overline W_1+ \alpha  X_3^o   \overline U_1  +   \beta\gamma  X_5^o   \overline U_1+ \alpha\gamma g  X_4   X_2^o  -   \sfrac12{g'}^2\kappa^2 \beta  X_4^w  X_4^u \right)\\
&&\mbox{}  -\alpha  s_2 X_7  \overline W_1  -  \sfrac14{g'}^2 \kappa^2  s_3 X_3^u   X_5^w - \alpha\beta\gamma(\bar\alpha^2 \bar s_0 + g\beta  s_1 )  X_4^o   X_6 .
\label{E08}
\EA
Additional constants are:
$$
\begin{tabular}{ l l l l l }
$\bar\kappa^2_\alpha=\alpha^2 +\bar p_0$, &\quad $\bar\kappa^2_\gamma = \gamma^2+\bar p_0$, &\quad $\bar\kappa^2_{\alpha\gamma} = \alpha^2+\gamma^2+\bar p_0$,&\quad
$ s_2 =  \bar s_0 + \beta s_1 $,&\quad $  s_3  = \bar s_0 + 2\beta  s_1$.
 \end{tabular}
$$

\subsection{Equations stemming from the dynamics of $\Psi_1$}
\BA
\NN\DDt X_4 + \nu \kappa^2_\alpha  X_4 &   =    & - \alpha  b   X_1 +   s_1 \left(\alpha\gamma (  X_2   X_5 + X_4^u   X_2^u -  X_1^w   X_3^w - X_2^o   X_1^o  ) - 2 \alpha(  X_1   \overline U_1 +  X_3^u   \overline U_0 )\right.\\
&&\hspace{4ex}\left.  \mbox{}-\beta\gamma^2 ( X_1^w   X_6^w  +  X_2   X_7 )\right)+  s_4 \left(\alpha  X_4^w   \overline W_0  -\sfrac12 \gamma^2(X_6   X_3 + X_5^w   X_2^w )\right),
\label{E09}\\
\NN \DDt  X_3^u  + \nu\kappa^2_\alpha   X_3^u &   =    & \alpha  b   X_1^u  +  s_1  \left(\alpha\gamma(   X_1^w   X_3^o + X_2^o   X_2^w   -  X_2   X_5^u  - X_4^u   X_3 ) + 2\alpha(  X_1^u   \overline U_1 + X_4   \overline U_0 ) \right.\\
&& \hspace{4ex} \left. \mbox{}-\beta\gamma^2 ( X_1^w   X_5^o  +  X_2   X_6^u )\right) -  s_4  \left(\alpha  X_4^o   \overline W_0  +\sfrac12\gamma^2(  X_5^w   X_1^o  +  X_6   X_2^u )\right),
\label{E10}\\
\NN \DDt  X_4^u  + \nu \kappa^2_\gamma  X_4^u  &  =    & s_1  \left(\alpha\gamma(  X_1   X_5^u + X_3^u   X_3   -  X_1^u   X_5- X_4   X_2^u ) + 2\gamma( X_1^w   \overline W_1 +  X_2^o   \overline W_0 )\right. \\
&& \hspace{4ex}\left. \mbox{} - \alpha^2\beta ( X_1^u   X_7 + X_1   X_6^u)\right) +  s_4  \left(\gamma  X_6   \overline U_0  - \sfrac12\alpha^2( X_4^o   X_2^w  +    X_4^w X_1^o )\right),
\label{E11}\\
\NN \DDt  X_2^o  + \nu\kappa^2_\gamma  X_2^o  &  =    &  s_1  \left( \alpha\gamma (  X_1^u   X_3^w + X_4   X_1^o -  X_1  X_3^o -  X_3^u   X_2^w ) - 2\gamma(  X_2   \overline W_1 +  X_4^u   \overline W_0 ) \right.\\
&& \hspace{4ex}\left.  \mbox{} -  \alpha^2\beta (  X_1^u   X_6^w + X_1  X_5^o )\right) -  s_4  \left(\gamma  X_5^w   \overline U_0 +\sfrac12\alpha^2( X_4^o   X_3 + X_4^w   X_2^u ) \right),
\label{E11}\\
\NN\DDt  X_5  + \nu\kappa^2_{\alpha\gamma}  X_5  &  =    & \alpha  b   X_3 
+  s_1  \left( 2\bar\alpha^2 \alpha (   X_3   \overline U_1+ X_5^u   \overline U_0) - 2\bar\gamma^2 \gamma(  X_1^o  \overline W_1+ X_3^w   \overline W_0) \right.\\
\NN && \hspace{4ex} \left. \mbox{}-2\alpha\gamma g(  X_1^u   X_4^u + X_4   X_2 )\vphantom{ X_1^o  \overline W_1 } \right) -  \sfrac14{g'}^2\kappa^2  s_4  \left( X_1 X_6  +  X_1^w X_4^w \right)\\
&&\hspace{4ex} \mbox{}  - \alpha(\bar s_1 + 2 \bar\alpha^2 \beta  s_1 )  X_6^w   \overline W_0  - \gamma(\bar s_1 + 2\bar\gamma^2 \beta  s_1 ) X_6^u   \overline U_0 ,
\label{E13}\\
\NN \DDt  X_5^u  + \nu\kappa^2_{\alpha\gamma}  X_5^u  &  =    & - \alpha  b   X_2^u 
-  s_1  \left(2\bar\alpha^2\alpha (  X_2^u   \overline U_1 + X_5   \overline U_0 )+ 2 \bar\gamma^2 \gamma(  X_2^w  \overline W_1 + X_3^o   \overline W_0 ) \right.\\
\NN && \hspace{4ex} \left.  \mbox{}+  2\alpha\gamma g( X_1   X_4^u + X_3^u   X_2 )\vphantom{ X_1^o  \overline W_1 } \right) -  \sfrac14{g'}^2\kappa^2  s_4  (  X_1^u   X_6 +  X_4^o  X_1^w ) \\
&&\hspace{4ex} \mbox{}+\alpha(\bar s_1 + 2 \bar\alpha^2 \beta  s_1 )  X_5^o   \overline W_0  - \gamma(\bar s_1 + 2\bar\gamma^2 \beta  s_1 )  X_7   \overline U_0 ,
\label{E14}\\
\NN \DDt  X_3^w  + \nu\kappa^2_{\alpha\gamma}   X_3^w  &  =    & \alpha  b   X_2^w 
+ s_1 \left(2\bar\alpha^2\alpha (  X_2^w   \overline U_1 + X_3^o   \overline U_0 )
+ 2\bar\gamma^2 \gamma ( X_2^u   \overline W_1  +  X_5   \overline W_0 )\right.\\
\NN && \hspace{4ex} \left.\mbox{}+ 2\alpha\gamma g( X_1^u   X_2^o  +  X_4   X_1^w )\vphantom{ X_1^o  \overline W_1 }\right)- \sfrac14{g'}^2\kappa^2  s_4  (  X_1   X_5^w +  X_4^w  X_2)\\
&& \hspace{4ex}\mbox{}
 -\alpha(\bar s_1 + 2 \bar\alpha^2 \beta  s_1 )  X_7   \overline W_0 + \gamma(\bar s_1 + 2\bar\gamma^2 \beta  s_1 )  X_5^o   \overline U_0
\label{E15}\\
\NN\DDt  X_3^o  + \nu\kappa^2_{\alpha\gamma}  X_3^o  &  =    & - \alpha  b   X_1^o 
-  s_1  \left( 2\bar\alpha^2\alpha( X_3^w   \overline U_0  +  X_1^o   \overline U_1 ) - 2\bar\gamma^2 \gamma\left( X_5^u   \overline W_0  +  X_3  \overline W_1 \right)\right.\\
\NN && \hspace{4ex} \left. \mbox{}+ 2\alpha\gamma g( X_1   X_2^o  +  X_3^u   X_1^w )\vphantom{ X_1^o  \overline W_1 }\right)- \sfrac14{g'}^2\kappa^2  s_4   \left(  X_1^u   X_5^w  +  X_4^o  X_2 \right)\\
&&\hspace{4ex}\mbox{}
 + \gamma(\bar s_1 + 2\bar\gamma^2 \beta  s_1 )  X_6^w   \overline U_0 + \alpha(\bar s_1 + 2 \bar\alpha^2 \beta  s_1 )  X_6^u   \overline W_0 ,
\label{E16}
\EA
Additional constants are:
$$
\begin{tabular}{ lll l }
$\kappa^2_\alpha=\alpha^2 +\bar p_1$, &\quad $\kappa^2_\gamma = \gamma^2+\bar p_1$, & \quad$\kappa^2_{\alpha\gamma} = \alpha^2+\gamma^2 +\bar p_1$, &\quad $ s_4 =     \bar s_1 + 2\beta  s_1 $.
\end{tabular}
$$

\subsection{Equations stemming from the dynamics of $\Phi_1$}
\BA
\NN&&\mu^2_\gamma \DDt   X_6  + \nu\kappa'^4_\gamma  X_6   =    \alpha^2 r   (   X_1^u   X_5^u  + X_4   X_3 +  X_1   X_5 + X_3^u   X_2^u  )\\ 
&&\hspace{4em}\mbox{}+d_\gamma \left( X_4^w   X_2^w  -  X_4^o   X_1^o \right) +  e_\gamma( X_1^u   X_6^u  -  X_1   X_7 )- c_\gamma  X_5^w   \overline W_0 ,
\label{E17}\\
\NN && \mu^2_\alpha \DDt  X_4^w  + \nu \kappa'^4_\alpha  X_4^w  =    \gamma^2 r    (X_1^w   X_5 +  X_2^o   X_2^u  +  X_2   X_3^w   + X_4^u   X_1^o) \\
&&\hspace{4em}\mbox{}+d_\alpha( X_5^w   X_3 - X_6  X_2^w )+e_\alpha(  X_1^w   X_7  - X_2   X_6^w ) + c_\alpha  X_4^o   \overline U_0 ,
\label{E18}\\
\NN&& \mu^2_\gamma\DDt  X_5^w  + \nu\kappa'^4_\gamma  X_5^w   =    \alpha^2 r    (   X_1^u   X_3^o  +  X_4   X_2^w +  X_1   X_3^w +  X_3^u   X_1^o )\\
 &&\hspace{4em}\mbox{}+ d_\gamma \left( X_4^o   X_2^u  -  X_4^w   X_3 \right)+ e_\gamma ( X_1   X_6^w  -  X_1^u   X_5^o ) +  c_\gamma  X_6   \overline W_0 ,
\label{E19}\\
\NN &&\mu^2_\alpha\DDt  X_4^o  + \nu \kappa'^4_\alpha X_4^o   =    
\gamma^2 r   ( X_1^w   X_5^u  +  X_2^o   X_3 +  X_2   X_3^o  + X_4^u   X_2^w  ) \\
 &&\hspace{4em}\mbox{}+ d_{\alpha} ( X_6   X_1^o  -  X_5^w   X_2^u )
  +e_\alpha ( X_2   X_5^o  -  X_1^w   X_6^u ) - c_\alpha X_4^w   \overline U_0  ,
\label{E20}\\
\NN && \mu^2_{\alpha\gamma}  \DDt  X_7  + \nu \kappa'^4_{\alpha\gamma}  X_7  =      \sfrac12 g'^2 \kappa^2 r    (  X_1^u  X_4^u +  X_4   X_2 )\\
&&\hspace{4em}\mbox{} + c_\kappa \left((\gamma X_6^w   \overline W_0 -\alpha  X_6^u   \overline U_0 )+   \sfrac12{g'}(\gamma^2 X_1   X_6 -\alpha^2 X_4^w   X_1^w )\right),
\label{E21}\\
\NN && \mu^2_{\alpha\gamma}   \DDt  X_6^u +\nu\kappa'^4_{\alpha\gamma}  X_6^u  =    \sfrac12 g'^2 \kappa^2 r    ( X_1   X_4^u + X_3^u   X_2 )\\
 &&\hspace{4em}\mbox{} +  c_\kappa \left((  \gamma  X_5^o  \overline W_0  + \alpha  X_7   \overline U_0 )      +   \sfrac12{g'}( \alpha^2  X_4^o   X_1^w  - \gamma^2  X_1^u   X_6 )\right).
\label{E22}\\
\NN && \mu^2_{\alpha\gamma} \DDt  X_6^w  +\nu\kappa'^4_{\alpha\gamma}  X_6^w  =        \sfrac12 g'^2 \kappa^2 r    ( X_1^u   X_2^o + X_4   X_1^w ) \\  
&&\hspace{4em}\mbox{} -  c_\kappa\left((\alpha   X_5^o   \overline U_0  +\gamma  X_7   \overline W_0 ) +   \sfrac12{g'} (\gamma^2 X_1   X_5^w   -\alpha^2   X_4^w   X_2 ) \right),
\label{E23}\\
\NN&&\mu^2_{\alpha\gamma} \DDt  X_5^o + \nu\kappa'^4_{\alpha\gamma} X_5^o   =      \sfrac12 g'^2 \kappa^2 r    (   X_1   X_2^o  +  X_3^u   X_1^w )\\
 &&\hspace{4em}\mbox{}  + c_\kappa\left((\alpha  X_6^w   \overline U_0 -\gamma  X_6^u   \overline W_0 ) +   \sfrac12{g'} (\gamma^2 X_1^u   X_5^w - \alpha^2  X_4^o   X_2 )\right).
\label{E24}
\EA
Additional constants are:
$$
\begin{tabular}{ lll }
$\mu^2_\alpha = {\alpha^2 + \beta^2}$, &\quad $\mu^2_\gamma = {\gamma^2 + \beta^2}$,&\quad $\mu^2_{\alpha\gamma} = \alpha^2+\gamma^2 + \beta^2$, \\[1ex]
$\kappa'^4_\alpha = \alpha^4 +2 \beta^2 \alpha^2 + p_1 $,&\quad $\kappa'^4_\gamma = \gamma^4 +2 \beta^2 \gamma^2 + p_1 $,&\quad $\kappa'^4_{\alpha\gamma} =\kappa^4 + 2 \beta^2 \kappa^2 + p_1$.\\[1ex]
$c_\alpha  =    \alpha( \bar r  + 2 \beta r    +\alpha^2 q  )$,&\quad $d_\alpha  =    \sfrac12 \gamma^2 \tau ( \bar r  + \alpha^2 q  )$,&\quad $e_\alpha  =     \sfrac12 \tau(\kappa^2 (\alpha^2 q  + \bar r ) + 2\alpha^2\beta r   )$,\\[1ex]
 $c_\gamma  =    \gamma( \bar r  + 2 \beta r    +\gamma^2 q  )$,&\quad $d_\gamma  =    \sfrac12\alpha^2\tau^{-1}(  \bar r +\gamma^2 q  )$,&\quad $e_\gamma  =     \sfrac12 \tau^{-1}(\kappa^2 (\gamma^2 q  + \bar r ) + 2\gamma^2\beta r   )$,\\[1ex]
$c_\kappa  =     ( \bar r  + 2\beta r    +\kappa^2 q  )$.
\end{tabular}
$$
In the stress-free for which $p_1=\beta^4$, one gets $\kappa'^4_\alpha=\mu^2_\alpha$,  $\kappa'^4_\gamma=\mu^2_\gamma$,  $\kappa'^4_{\alpha\gamma}=\mu^2_{\alpha\gamma}$.

\subsection{Equations governing mean flow amplitudes $\overline{\mathcal U}$}
\BA
\DDt  \overline U_0 + \nu \bar p_0  \overline U_0 = \sfrac14\gamma \bar s_0[2\gamma^2(  X_6 X_4^u - X_2^o X_5^w )
+ \kappa^2 (   X_5 X_6^u  +  X_5^u X_7 -  X_5^o X_3^w - X_6^w X_3^o )],
\label{E25}\\
\DDt \overline W_0 +\nu \bar p_0  \overline W_0 = \sfrac14 \alpha \bar s_0 [ 2\alpha^2( X_4  X_4^w - X_4^o X_3^u )+ \kappa^2 (  X_5^o  X_5^u  + X_3^o  X_6^u  - X_5 X_6^w - X_7 X_3^w)],
\label{E26}\\
\DDt  \overline U_1 + \nu \bar p_1  \overline U_1 = \sfrac14\gamma \bar s_1[2\gamma^2( X_6 X_2 - X_1^w X_5^w )  +  \kappa^2 ( X_6^w X_2^w + X_5^o X_1^o  - X_2^u X_6^u  - X_3 X_7 )],
\label{E27}\\
\DDt {\overline W}_1 + \nu \bar p_1 {\overline W}_1 = \sfrac14\alpha \bar s_1 (2\alpha^2 ( X_1^u X_4^w - X_4^o X_1 ) + \kappa^2 ( X_5^o X_3 + X_2^w  X_6^u - X_2^u X_6^w - X_7 X_1^o )].
\label{E28}
\EA

\bibliography{bibliography}